\documentclass[12pt, draftclsnofoot, onecolumn]{IEEEtran}
\usepackage{amsmath,amsfonts,color}
\usepackage{algorithmic}
\usepackage{algorithm}
\usepackage{array}
\usepackage{textcomp}
\usepackage{stfloats}
\usepackage{url}
\usepackage{verbatim}
\usepackage{graphicx}
\usepackage[caption=false,font=footnotesize,labelfont=rm,textfont=rm]{subfig}
\usepackage{cite}
\usepackage{cuted}
\usepackage{setspace}

\newtheorem{proposition}{Proposition}
\newtheorem{remark}{Remark}

\begin{document}

\title{OTFS-based Robust MMSE Precoding Design in Over-the-air Computation}

\author{Dongkai Zhou, Jing Guo, \IEEEmembership{Senior Member,~IEEE,} Siqiang Wang, Zhong Zheng, \IEEEmembership{Member,~IEEE,} 
	 Zesong Fei, \IEEEmembership{Senior Member,~IEEE}, Weijie Yuan, \IEEEmembership{Member,~IEEE}, and Xinyi Wang, \IEEEmembership{Member,~IEEE}
\thanks{Dongkai Zhou, Jing Guo, Siqiang Wang, Zhong Zheng, Zesong Fei, and Xinyi Wang are with the School of Information and Electronics, Beijing Institute of Technology, Beijing 100081, China (e-mail: \{3120220778, jingguo, 3120205406, zhong.zheng, feizesong, wangxinyi\}@bit.edu.cn). Weijie Yuan is with the Department of EEE, Southern University of Science and Technology, Shenzhen 518055, China (e-mail: yuanwj@sustech.edu.cn).}
}



\maketitle

\begin{abstract}
Over-the-air computation (AirComp), as a data aggregation method that can improve network efficiency by exploiting the superposition characteristics of wireless channels, has received much attention recently. Meanwhile, the orthogonal time frequency space (OTFS) modulation can provide a strong Doppler resilience and facilitate reliable transmission for high-mobility communications. Hence, in this work, we investigate an OTFS-based AirComp system in the presence of time-frequency dual-selective channels. In particular, we commence from the development of a novel transmission framework for the considered system, where the pilot signal is sent together with data, and the channel estimation is implemented according to the echo from the access point to the sensor, thereby reducing the overhead of channel state information (CSI) feedback. Hereafter, based on the CSI estimated from the previous frame, a robust precoding matrix aiming at minimizing mean square error in the current frame is designed, which takes into account the estimation error from the receiver noise and the outdated CSI. The simulation results demonstrate the effectiveness of the proposed robust precoding scheme by comparing it with the non-robust precoding. The performance gain is more obvious in a high signal-to-noise ratio in case of large channel estimation errors.
\end{abstract}

\begin{IEEEkeywords}
Over-the-air computation, orthogonal time frequency space, imperfect channel state information, robust precoding.
\end{IEEEkeywords}

\section{Introduction}
The Internet of Everything is an important application scenario in the future 6G communication systems, which generally requires huge spectrum resources \cite{b1}. Over-the-air computation (AirComp) is regarded as a promising solution to the problem of limited spectrum resources \cite{b2}. The AirComp technology allows the concurrent transmission of multiple nodes. Rather than treating the signals from other nodes as noise, it leverages the signal superposition property of co-channels to compute a class of nomographic functions, e.g., weighted sum and arithmetic mean, of the distributed sensing data, whereby improving the efficiency of the wireless communication system. 

The idea of AirComp first came from the study on the computation functions in multiple-access channels in \cite{b3}. Later on, there is much literature investigating the AirComp system from the perspective of signal alignment \cite{b4}, power control \cite{b5}, beamforming design \cite{b00}, etc., under the assumption of perfect channel state information (CSI). In practice, the channel estimation may not be perfect due to the factors such as noise. Hence, some other works designed the AirComp transmission scheme with the inclusion of a channel estimation procedure. Specifically, the authors in \cite{b6} proposed a two-stage architecture, where in the first stage the fusion center obtained the sum channel gain according to the reference signal from sensors, and the estimated CSI was used in the second stage for data transmission. Based on the same architecture, in \cite{b7}, the sensors utilized pilot signals broadcast by the fusion center to obtain local CSI, and a low overhead CSI feedback algorithm was designed. In \cite{b8}, the impact of imperfect CSI on the computation accuracy of AirComp was studied, and a transceiver design on the basis of the statistical error of channel estimation was developed. 

The channel estimation and data transmission in aforementioned works \cite{b6,b7,b8} happened in different phases, which can incur the signaling overhead. Besides, existing works designed the transmission mechanism based on orthogonal frequency division multiplexing waveform or investigated the AirComp system in a static scenario, where the impact of the multipath effect and Doppler shift were not considered. Note that for the high-mobility scenarios (e.g., the fusion center is vehicular or drone), the channel becomes a time-frequency doubly-selective channel, which makes the schemes in the literature fail to work.

For reliable communications over time-frequency doubly-selective channels, orthogonal time frequency space (OTFS), a recently proposed two-dimensional (2D) multi-carrier modulation technique, is a promising candidate \cite{b04,b05}. OTFS modulates the information symbols in the Delay-Doppler (DD) domain and each symbol is mapped to the entire time-frequency (TF) domain by 2D transformation, which takes advantage of the full TF diversity \cite{b06}. Additionally, it converts the complex time-varying channel in the TF domain into a sparse and stable channel in the DD domain \cite{b07}, which helps to perform better channel estimation and equalization. To the best knowledge of the authors, the application of the OTFS signaling to the AirComp system has not been investigated in the literature yet.

Inspired by the above discussions, in this paper, we propose an AirComp system based on OTFS waveform, which contains multiple sensors with dual functions of radar and communications and an unmanned aerial vehicle (UAV) as an access point (AP). More specifically, with the advantages of dual-function sensors, we first come up with a novel transmission scheme together with the frame structure. In this scheme, the estimation of CSI no longer occupies a separate phase. Instead, the sensor uses the echo from the AP to assist the CSI estimation. Such implementations can greatly reduce the signaling overhead and improve the system’s efficiency. The estimated CSI in the current frame is utilized to design a precoding matrix for the next frame to eliminate the effect of the time-frequency doubly-selective channel. Hence, by taking into account the errors in the estimated CSI and the error caused by the outdated CSI, we then propose a robust precoding design relying on the statistical characteristics of errors. Our numerical results demonstrate that our developed robust precoder outperforms the non-robust precoder, especially in a high signal-to-noise ratio (SNR) scenario, which indicates the importance of the inclusion of imperfect CSI.

\section{System Model}
\subsection{Network Model}
Let us consider a data aggregation scenario for a wireless sensor network, which is composed of $Q$ sensors and a UAV acting as the AP. Both the sensors and the UAV are assumed to be equipped with a single antenna. The sensors residing in a certain region sense the environment information and transmit it to the UAV, while the UAV hovering in this region aggregates and processes the sensing data, e.g., arithmetic mean. Since the computation capability of the UAV is relatively weak, the UAV is assumed to implement data aggregation via the AirComp technology, thereby avoiding the complicated signal processing process at the UAV. Moreover, similar to \cite{b4,b5,b00}, symbol-level synchronization is assumed. In this work, to eliminate the influence of high mobility channels, the transmission between each sensor and the UAV is carried out in the delay-Doppler domain, i.e., the OTFS waveform is exploited. 

\subsection{Proposed OTFS-based Transmission Framework}

\begin{figure}[!t]
	\centerline{\includegraphics[width=0.9\linewidth]{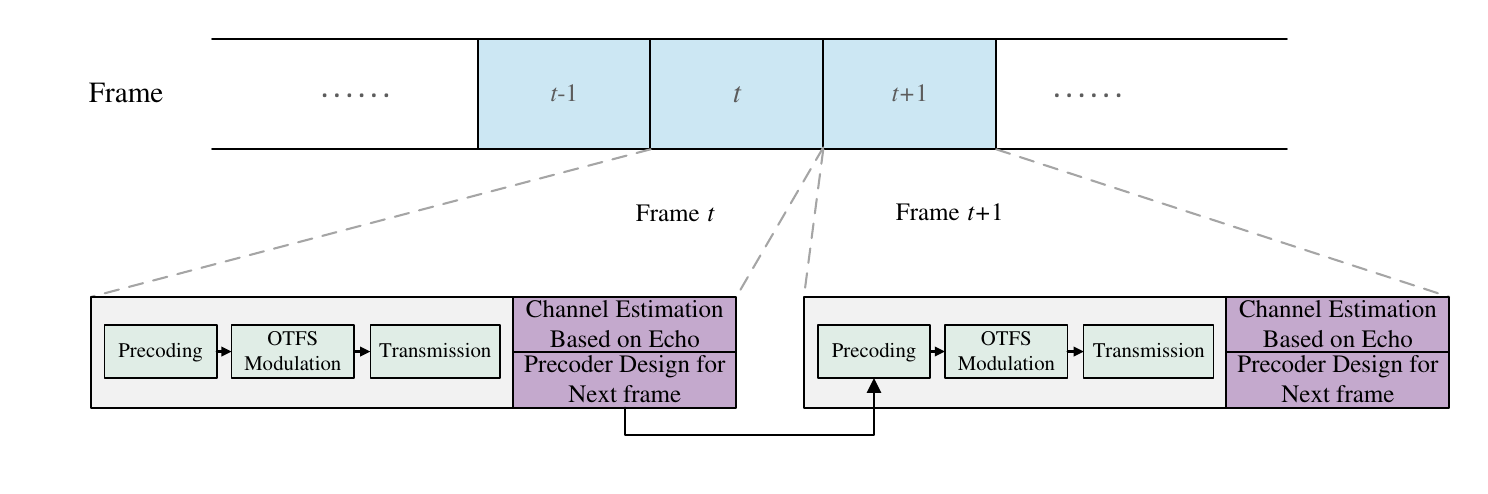}}
	\caption{The illustration of the transmission framework of the sensor in the considered system.}
	\label{fig1}
\end{figure}

For the existing works on AirComp, e.g., \cite{b5,b00,b7}, the CSI between the AP and each node is estimated by the sensor node based on the pilot signal broadcast by the AP. Then the CSI is fed back to the AP through the sensor node. Such a transmission mechanism can easily result in the signaling overhead for CSI exchange. Hence, to reduce the signaling overhead and the transmission of the UAV, we develop a simplified OTFS-based transmission framework for our considered system setup which integrates the channel estimation with the AirComp by reusing the pilots transmitted by sensor nodes for their channel estimation. As illustrated in Fig.~\ref{fig1}, the proposed framework contains two procedures during each frame, i.e., 
\begin{itemize}
	\item At the first stage, each sensor performs OTFS modulation including precoding and transmits the signal to the AP.   
	\item At the second stage, each sensor estimates the CSI based on the echo from the AP, and the recovered channel is then used to design a precoding matrix for the next frame. 
\end{itemize}

\begin{figure}[!t]
	\centerline{\includegraphics[width=0.7\linewidth]{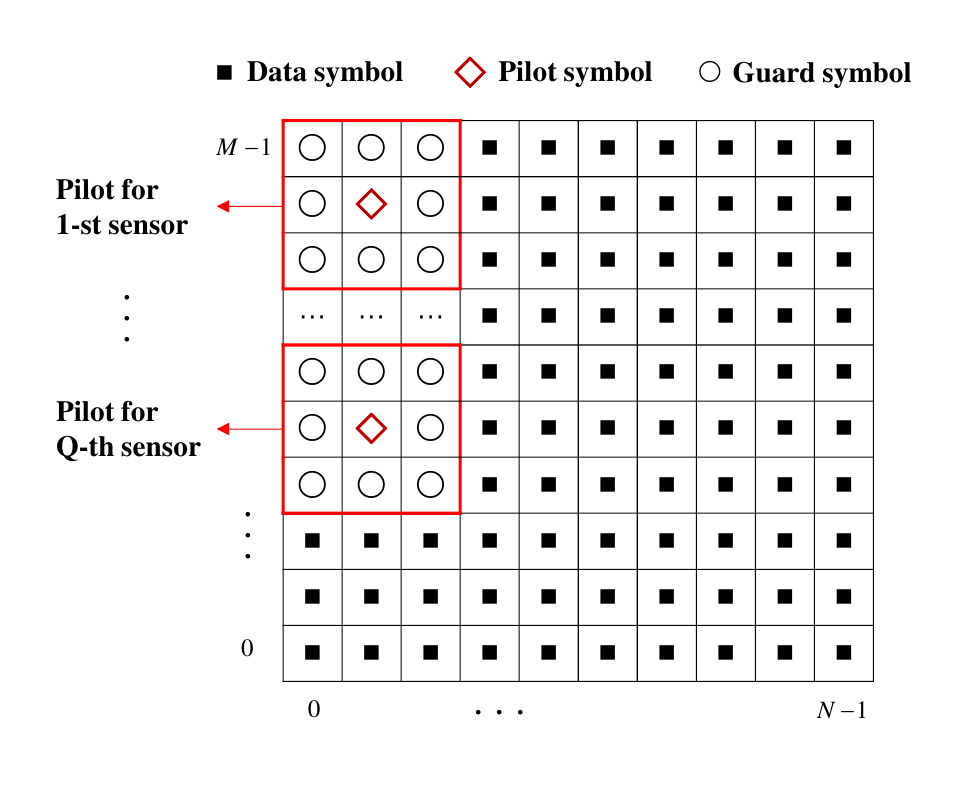}}
	\caption{The schematic diagram of the frame structure of the OTFS-based  signal.}
	\label{fig2}
\end{figure}

The frame structure for the proposed framework is depicted in Fig.~\ref{fig2}. Therein, the data symbols of the sensors are all arranged in the same position. The pilot of each sensor is placed in different positions on the resource grid, which can eliminate the interference coming from other sensors during channel estimation\footnote{In this work, we assume that the number of sensors is not very large such that the position of the pilot for each sensor is orthogonal. The consideration of non-orthogonal placement is left for our future work. }.

The detailed signal models are described below. Let $\mathbf{X}_q \in \mathbb{C}^{M \times N} (q=1,2,..., Q) $ denote the data sent by the $q$-th sensor. Via vectorization, the transmit data in vector form can be expressed as $\mathbf{x}_q=\text{vec} (\mathbf{X}_q) \in \mathbb{C}^{MN \times 1}$. By applying a precoder $\mathbf{F}_q \in \mathbb{C}^{MN \times MN}$ in the DD domain, the transmit signal $\mathbf{d}_q \in \mathbb{C}^{MN \times 1}$ of the $q$-th sensor can be expressed as
%
\begin{equation}
	\mathbf{d}_q = \mathbf{F}_q \mathbf{x}_q. \label{eq1}
\end{equation}
After the process of OTFS modulation with a rectangular pulse shaping filter \cite{b08}, $\mathbf{d}_q$ is converted into the transmit signal in the time domain, denoted as $\mathbf{s}_q \in \mathbb{C}^{MN \times 1}$, which can be obtained by
\vspace{-0.1cm}
\begin{equation}
	\mathbf{s}_q = \left( \mathbf{W}_{N}^\text{H}\otimes {{\mathbf{I}}_{M}} \right) \mathbf{d}_q, \label{eq2}
\end{equation}
where $\mathbf{W}_{N}^\text{H}$ is the inverse discrete Fourier transform matrix of order $N$ and $\mathbf{I}_{M}$ denotes the identity matrix of size $M \times M$. 

The time domain channel matrix between the $q$-th sensor and AP is defined as $\mathbf{H}^{\text{TD}}_{q} \in \mathbb{C}^{MN \times MN}$. Note that the channel among different sensors can be either correlated or independent. According to \cite{b08}, $\mathbf{H}^{\text{TD}}_{q}$ can be expressed as
\vspace{-0.1cm}
\begin{equation}
	\mathbf{H}^{\text{TD}}_{q} =\sum_{p=1}^{P} h_{p,q} \mathbf{\Pi}^{l_{p,q}} \mathbf{\Delta}^{k_{p,q}}, \label{eq3}
	\vspace{-0.1cm}
\end{equation}
where $P$ is the number of resolvable paths between the sensor and AP. $h_{p,q}\sim \mathcal{CN}(0,\frac{1}{P})$ is the channel gain of the $p$-th path; ${l}_{p,q}$ and ${k}_{p,q}$ denote the delay taps and Doppler taps at the $p$-th path, respectively.
${\mathbf{\Pi }}$ is the permutation matrix characterizing the delay influence, expressed as $\mathbf{\Pi}=\text{circ}\left\{ { [0,1,...,0]^{\text{T}}_{MN \times 1} }\right\}$, where $\text{circ}\{\mathbf{x}\}$ represents the matrix consisting of cyclic shifts of $\mathbf{x}$ and
${\mathbf{\Delta }}$ is a diagonal matrix characterizing the Doppler influence, which is defined as $\mathbf{\Delta }=\text{diag}\left\{ {{\left[ {{e}^{\frac{j2\pi }{MN}\times 0}},{{e}^{\frac{j2\pi }{MN}\times 1}},...,{{e}^{\frac{j2\pi }{MN}\times \left( MN-1 \right)}} \right]}^{\text{T}}} \right\}$, where $\text{diag}\{\mathbf{x}\}$ represents the matrix with $\mathbf{x}$ being its diagonal elements.

Due to the wave-addition of the multi-access channel \cite{b00}, the received signal $\mathbf{y} \in \mathbb{C}^{MN \times 1}$ at the AP can represented as
\begin{align}
	\mathbf{y} &= \sum_{q=1}^{Q}\frac{\left(\gamma_q \mathbf{H}^{\text{TD}}_{q} \mathbf{s}_q+\mathbf{n}_q \right)}{\gamma_q} \nonumber
	\\ & = \sum_{q=1}^{Q}\left(\mathbf{H}^{\text{TD}}_{q} \left( \mathbf{W}_{N}^{\text{H}}\otimes {{\mathbf{I}}_{M}} \right)\mathbf{F}_q \mathbf{x}_q+\frac{\mathbf{n}_q}{\gamma_q} \right) \nonumber
	\\ & = \sum_{q=1}^{Q}\left( \mathbf{H}_{q} \mathbf{F}_q \mathbf{x}_q+\frac{\mathbf{n}_q}{\gamma_q} \right), \label{eq4}
\end{align}
where $\mathbf{H}_{q} \in \mathbb{C}^{MN \times MN}$ is the equivalent channel matrix of the $q$-th sensor and $\mathbf{n}_q$ is the additive white Gaussian noise (AWGN), satisfying $\mathbb{E}\left[\mathbf{n}_q \mathbf{n}_q^\text{H} \right]=\sigma_{n}^2\mathbf{I}$. $\gamma_q$ is the power normalization factor that keeps the sensor transmit power constant, which is denoted as $\gamma_q=\sqrt{\frac{P_t}{\text{trace}\left(\mathbf{F}_q\mathbf{F}^\text{H}_q\right)}}$. $P_t$ is the total power of the transmit data symbol that is assumed to be the same for all sensors.

\section{Robust Precoding Design for OTFS}
In this work, taking the sum function of all sensor data as the computation target, we aim to design a robust precoder based on the estimated CSI for our proposed transmission framework. The channel estimation, the corresponding estimation error modeling, and the precoding design are presented in the following.

\subsection{Channel Estimation and Error Modeling}
As described in Section II, each sensor estimates the CSI according to the echo from the AP. During this stage, to obtain the estimated value of round-trip delay tap $\tilde{l}_p$, Doppler tap $\tilde{k}_p$ and channel gain $\tilde{h}_p \ (p=1,2,...,P)$, the single pilot-based estimation method in \cite{bx} is adopted\footnote{The focus of this work is not to propose a channel estimation approach. Hence, we adopt the widely considered single pilot-based scheme. Note that, in the case of the orthogonal arrangement of pilots, the single pilot-based scheme in \cite{bx} incurs smaller pilot overhead compared to the multiple pilot-based scheme in \cite{b06}, making it more suitable for our proposed system. Furthermore, although the superimposed pilot-based transmission scheme developed in \cite{SP1,SP2} can improve the spectral efficiency by superimposing the high-power pilot symbols on data symbols, this approach requires complex algorithms at the receiver to eliminate the interference between the pilot and data. This does not apply to our proposed transmission framework, as we assume that the AP has limited capabilities to perform data processing and instead directly treats the received signals as the computation results.}, where the subscript $q$ of
the sensor is omitted for ease of illustration. We assume that the values of the round trip delay taps and Doppler taps are twice as large as those of the one-way communication channel from the sensor to AP, and the channel gains of both taps are the same, that is, $\tilde{l}_p = 2{l}_p$, $\tilde{k}_p = 2{k}_p$, and $\tilde{h}_p = h_p$ \cite{bxx}. Thus, the estimated value can be obtained by $\hat{l}_p=\tilde{l}_p/2$, $\hat{k}_p=\tilde{k}_p/2$, and $\hat{l}_p=\tilde{h}_p$, respectively.

Due to the random factors such as receiver noise, the channel estimation may not be perfect. In this work, we mainly focus on the estimation error occurred on the channel gain. In terms of the delay taps and Doppler taps, under the proper pilot setup, the estimation can be regarded as accurate \cite{bx}.
Note that, in our result section, we will show the robustness of our precoding design by including the impacts of the estimation error of delay taps and Doppler taps. 
The channel estimation error for the channel gain comes from two factors, i.e., the estimation error due to receiver noise and the outdated estimation error. The latter part comes from the inherent mechanism of our proposed transmission framework. Under our framework, the precoding matrix design in the current frame is based on the estimated CSI from the previous frame. The channel gains in two time frames are not exactly the same but highly correlated, which consequently causes the problem of outdated CSI. The error modeling for these two factors is demonstrated below.

For the estimation error due to receiver noise, according to \cite{bx}, the channel gain estimation result at the $(t-1)$-th frame via the pilot-based method is given by
\begin{equation}
	\hat{h}_{p,t-1} = \frac{h_{p,t-1} \theta_p x_{o} +w_p}{x_o\theta_p} = h_{p,t-1}+\frac{w_p}{x_o \theta_p}, \label{eq7}
\end{equation}
where $x_o$ is the pilot symbol, $\theta_p$ is a phase term associated with the pilot position and $w_p\sim \mathcal{CN}(0,\sigma^2_w)$ is the complex Gaussian noise at the sensor. The subscript $t$ is added to distinguish different frames. From \eqref{eq7}, $\hat{h}_{p,t-1}$ is also a complex Gaussian variable, satisfying $\hat{h}_{p,t-1}\sim \mathcal{CN}(0,\frac{1}{P}+\frac{\sigma^2_w}{x^2_o})$.

As for the outdated CSI, akin to \cite{bxxx}, the relationship between $h_{p,t}$ and $h_{p,t-1}$ is characterized by
\begin{equation}
	h_{p,t} = \rho h_{p,t-1} + \sqrt{1-\rho^2}z_p,\label{eq8}
\end{equation}
where $\rho \in(0,1) $ is the correlation coefficient, and $z_p \sim \mathcal{CN}(0,\frac{1}{P})$ is a complex Gaussian noise. Assuming that the first term of the correlation coefficient in \eqref{eq8} can be compensated, the exact CSI at the $t$-th frame $h_{p,t}$ is related to the estimated CSI at the $(t-1)$-th frame $\hat{h}_{p,t-1}$ by
\begin{equation}
	{h}_{p,t} = \rho \hat{h}_{p,t-1} + e_p, \label{eq9}
\end{equation}
where $e_p\sim \mathcal{CN}(0,\sigma^2_e)$ and $\sigma^2_e = \rho^2 \frac{\sigma^2_w}{x^2_o} + \frac{(1-\rho^2)}{P}$.

Bringing \eqref{eq9} back to \eqref{eq3}, we have the exact channel matrix $\mathbf{H}_q$ at the $t$-th frame related to the estimated channel matrix  $\hat{\mathbf{H}}_q$ recovered by $\rho \hat{h}_{p,t-1}$ at the $(t-1)$-th frame written as
\begin{subequations}
	\begin{align}
		\mathbf{H}_q &= \hat{\mathbf{H}}_q + \mathbf{E}_q, \label{eq10a}\\
		\mathbf{E}_q &= \mathbf{E}^{\text{TD}}_{q} \left( \mathbf{W}_{N}^\text{H}\otimes {{\mathbf{I}}_{M}} \right), \label{eq10b} \\
		\mathbf{E}^{\text{TD}}_{q} &=\sum_{p=1}^{P} e_{p,q} \mathbf{\Pi}^{l_{p,q}} \mathbf{\Delta}^{k_{p,q}}, \label{eq10c}
	\end{align}
\end{subequations}
where $\mathbf{E}_q$ is the error in the channel matrix, and $\mathbf{E}^{\text{TD}}_{q}$ is the error in the time domain channel matrix.

\subsection{Robust MMSE Precoder for OTFS}
For AirComp, to analyze the accuracy of the computation, the MSE between the true value of the target and the aggregation value is generally adopted as the performance metric, mathematically
\begin{align}
	\text{MSE}&=\mathbb{E}\left[\left| \mathbf{y}-\sum_{q=1}^{Q}\mathbf{x}_q \right|^2\right] \nonumber
	\\ &=\mathbb{E}\left[\sum_{q=1}^{Q}\left| \left(\mathbf{H}_{q} \mathbf{F}_q-\mathbf{I}\right)\mathbf{x}_q +\frac{\mathbf{n}_q}{\gamma_q}\right|^2 \right ]. \label{eq11}
\end{align}

In this work, we aim at minimizing the MSE by designing the precoding matrix $\mathbf{F}_q(q=1,...,Q)$. Since each sensor is independent of the others, we can separate the joint optimization problem into $Q$ independent problems, and the precoding matrix of each sensor shares the same closed-form solution. Furthermore, under the consideration of imperfect CSI, the channel matrix $\mathbf{H}_{q}$ in \eqref{eq11} needs to be replaced by $\hat{\mathbf{H}}_q + \mathbf{E}_q$. Therefore, for the $q$-th sensor, the MSE is written as
\begin{align}
	\text{MSE}_q =\mathbb{E}&\left[ \left| \left((\hat{\mathbf{H}}_{q}+\mathbf{E}_q) \mathbf{F}_q-\mathbf{I}\right)\mathbf{x}_q +\frac{\mathbf{n}_q}{\gamma_q} \right|^2\right ] \nonumber
	\\ =\mathbb{E}&\left[ \text{tr}\left(\left(\left((\hat{\mathbf{H}}_{q}+\mathbf{E}_q) \mathbf{F}_q-\mathbf{I}\right)\mathbf{x}_q +\frac{\mathbf{n}_q}{\gamma_q}\right) \right.\right. \nonumber \\ & \left.\left.\left(\left((\hat{\mathbf{H}}_{q}+\mathbf{E}_q) \mathbf{F}_q-\mathbf{I}\right)\mathbf{x}_q +\frac{\mathbf{n}_q}{\gamma_q}\right)^\text{H}\right)\right]. \label{eq12}
\end{align}


According to \eqref{eq12}, the optimization problem for the $q$-th sensor can be described as
\begin{equation}
	\underset{{{\mathbf{F}}_{q}}}{\mathop{\text{min}}}\,\quad \text{MSE}_{q}. \label{eq13}
\end{equation}

Since there is no constraint condition for the above optimization problem, the closed-form solution of $\mathbf{F}_q$ can be achieved by derivative, which is presented in the following proposition. 
\begin{proposition}
	For our proposed transmission framework of the OTFS-based AirComp system, the robust MMSE precoder of the $q$-th sensor at the current frame is given by
\begin{equation}
	\mathbf{F}_{q}^{*} = {\left( {{{\mathbf{\hat{H}}}^\text{H}_q}{\mathbf{\hat{H}}}_q + \left( {{{\sigma _n^2}} + P\sigma _e^2} \right){\mathbf{I}}} \right)^{ - 1}}{{\bf{\hat{H}}}^\text{H}_q}. \label{opt} 
\end{equation}
\end{proposition}

\begin{IEEEproof}
Under the assumption that data symbols are independently and identically distributed (i.i.d.) with zero mean and normalized variance, and the data and the noise are statistically independent, i.e., $\mathbb{E}\left[\mathbf{x}_q\mathbf{x}_q^\text{H}\right]=\mathbf{I} $, $\mathbb{E}\left[\mathbf{n}_q\mathbf{n}_q^\text{H}\right]=\sigma^2_n\mathbf{I} $, and $\mathbb{E}\left[\mathbf{x}_q\mathbf{n}_q^\text{H}\right]=\mathbf{0} $, the $\text{MSE}_q$ in \eqref{eq12} can be further simplified to \eqref{eq14} as follows
\begin{align}
	\text{MSE}_q &=\mathbb{E}\left[ \text{tr}\left( \left(\mathbf{\hat{H}}_{q} \mathbf{F}_{q}+\mathbf{E}_q\mathbf{F}_{q}- \mathbf{I}\right) \left(\mathbf{F}^\text{H}_{q}\mathbf{\hat{H}}^\text{H}_{q}+\mathbf{F}^\text{H}_{q}\mathbf{E}^\text{H}_q-\mathbf{I}\right) 
	+\sigma^2_n\mathbf{F}_{q}\mathbf{F}^\text{H}_{q}  \right)\right]\nonumber\\
	&=\text{tr}\left({{\mathbf{\hat{H}}}_{q}}{{\mathbf{F}}_{q}}\mathbf{F}_{q}^{\text{H}}\mathbf{\hat{H}}_{q}^{\text{H}}-{{\mathbf{\hat{H}}}_{q}}{{\mathbf{F}}_{q}}-\mathbf{F}_{q}^{\text{H}}\mathbf{\hat{H}}_{q}^{\text{H}}+\mathbf{I}+\sigma _{n}^{2}{{\mathbf{F}}_{q}}\mathbf{F}_{q}^{\text{H}} \right.\nonumber \\& \left.
	\quad \quad \ +\mathbb{E}\left[ {{\mathbf{\hat{H}}}_{q}}{{\mathbf{F}}_{q}}\mathbf{F}_{q}^{\text{H}}\mathbf{E}_{q}^{\text{H}}+{{\mathbf{E}}_{q}}{{\mathbf{F}}_{q}}\mathbf{F}_{q}^{\text{H}}\mathbf{\hat{H}}_{q}^{\text{H}}+{{\mathbf{E}}_{q}}{{\mathbf{F}}_{q}}\mathbf{F}_{q}^{\text{H}}\mathbf{E}_{q}^{\text{H}}-{{\mathbf{E}}_{q}}{{\mathbf{F}}_{q}}-\mathbf{F}_{q}^{\text{H}}\mathbf{E}_{q}^{\text{H}} \right] \right). \label{eq14}
\end{align}

Since the mean value of $e_p$ is 0, combined with \eqref{eq10a} - \eqref{eq10c}, we can obtain that the mean value of every element in $\mathbf{E}_{q}$ is 0. That is to say, $\mathbb{E}\left[\mathbf{E}_{q}\right] =\mathbb{E}\left[\mathbf{E}^\text{H}_{q}\right]=\mathbf{0} $. Thus, \eqref{eq14} can be further simplified as
\begin{equation}
		\text{MSE}_q =  \text{tr}\left({{\mathbf{\hat{H}}}_{q}}{{\mathbf{F}}_{q}}\mathbf{F}_{q}^{\text{H}}\mathbf{\hat{H}}_{q}^{\text{H}}-{{\mathbf{\hat{H}}}_{q}}{{\mathbf{F}}_{q}}-\mathbf{F}_{q}^{\text{H}}\mathbf{\hat{H}}_{q}^{\text{H}}+\mathbf{I}
		+\sigma _{n}^{2}{{\mathbf{F}}_{q}}\mathbf{F}_{q}^{\text{H}}+\mathbb{E}\left[ {\mathbf{E}_{q}^{\text{H}}\mathbf{E}_{q}} \right]{{\mathbf{F}}_{q}}\mathbf{F}_{q}^{\text{H}}  \right).\label{eq16}
\end{equation}

We then target at obtaining the exact expression of $\mathbb{E}\left[ {\mathbf{E}_{q}^{\text{H}}\mathbf{E}_{q}} \right]$. From \eqref{eq10b}, $\mathbb{E}\left[ {\mathbf{E}_{q}^{\text{H}}\mathbf{E}_{q}} \right]$ can be written as
\begin{equation}
	\mathbb{E}\left[ {\mathbf{E}_{q}^{\text{H}}\mathbf{E}_{q}} \right] = \left( \mathbf{W}_{N}^\text{H}\otimes {{\mathbf{I}}_{M}} \right)^{\text{H}}\mathbb{E}\left[{\mathbf{E}^{\text{TD}}_{q}}^{\text{H}} \mathbf{E}^{\text{TD}}_{q}\right]  \mathbf{W}_{N}^\text{H}\otimes {{\mathbf{I}}_{M}}. \label{eq17}
\end{equation}

Since $\mathbf{W}_{N}^\text{H}\otimes {{\mathbf{I}}_{M}}$ is a deterministic matrix, the next step is to calculate $\mathbb{E}\left[{\mathbf{E}^{\text{TD}}_{q}}^{\text{H}} \mathbf{E}^{\text{TD}}_{q}\right]$. According to the expression for $\mathbf{E}^{\text{TD}}_{q}$ in \eqref{eq10c}, $\mathbf{E}^{\text{TD}}_{q}$ is sparse and has non-zero values only on the diagonal and a few cyclic shifts of the diagonal. It can be expressed as follows
\begin{equation}
	\mathbf{E}^{\text{TD}}_{q} = \left[ \begin{matrix}
		{{e}_{1}}{{\alpha }_{1,1}} & 0 & \ldots  & \ldots  & \ldots  & 0  \\
		0 & {{e}_{1}}{{\alpha }_{1,2}} & \ddots  & \ddots  & \ddots  & \vdots   \\
		\vdots  & 0 & \ddots  & \ddots  & \ddots  & {{e}_{P}}{{\alpha }_{P,MN}}  \\
		{{e}_{P}}{{\alpha }_{P,1}} & \ddots  & \ddots  & \ddots  & \ddots  & \vdots   \\
		0 & {{e}_{P}}{{\alpha }_{P,2}} & \ddots  & \ddots  & \ddots  & \vdots   \\
		\vdots  & \ldots  & \ldots  & \ldots  & \ldots  & {{e}_{1}}{{\alpha }_{1,MN}}  \\
	\end{matrix} \right], \label{eq18}
\end{equation}
where $\alpha_{m,n} = e^{\frac{j2\pi}{MN}\left(n-1\right)k_{m,q}}\left(m=1,...,P,\ n=1,...,MN\right)$ is the phase term caused by Doppler. Let  $r_{mn}$ denote the $(m,n)$-th element in the matrix ${\mathbf{E}^{\text{TD}}_{q}}^{\text{H}} \mathbf{E}^{\text{TD}}_{q}$. Since $e_p$ is i.i.d., the mean value of $r_{mn}$ can be obtained as
\begin{align}
	\mathbb{E}\left[ {{r_{mn}}} \right] = \left\{ \begin{array}{l}
		P\sigma _e^2,\quad m = n\\
		0,{\kern 1pt} \ \quad\quad m \ne n
	\end{array} \right.. \label{eq19}
\end{align}
According to \eqref{eq19}, we can obtain $\mathbb{E}\left[{\mathbf{E}^{\text{TD}}_{q}}^{\text{H}} \mathbf{E}^{\text{TD}}_{q}\right]$ as
\begin{equation}
	\mathbb{E}\left[{\mathbf{E}^{\text{TD}}_{q}}^{\text{H}} \mathbf{E}^{\text{TD}}_{q}\right] = P\sigma _e^2 \ \mathbf{I}_{MN}.\label{eq20}
\end{equation}
Bringing \eqref{eq20} back to \eqref{eq17}, the term $\mathbb{E}\left[ {\mathbf{E}_{q}^{\text{H}}\mathbf{E}_{q}} \right]$ can be further simplified into
\begin{align}
		\mathbb{E}\left[ {\mathbf{E}_{q}^{\text{H}}\mathbf{E}_{q}} \right] &= P\sigma _e^2 \left( \mathbf{W}_{N}^\text{H}\otimes {{\mathbf{I}}_{M}} \right)^{\text{H}}  \mathbf{W}_{N}^\text{H}\otimes {{\mathbf{I}}_{M}} \nonumber \\
		&= P\sigma _e^2 \ \mathbf{I}_{MN}. \label{eq21}
\end{align}

Finally, by substituting \eqref{eq21} into \eqref{eq16}, we can obtain the expression for $\text{MSE}_q$ in \eqref{eq16} that exploits the statistical properties of the channel estimation error, expressed as
\begin{align}
		\text{MSE}_q =  \text{tr}&\left({{\mathbf{\hat{H}}}_{q}}{{\mathbf{F}}_{q}}\mathbf{F}_{q}^{\text{H}}\mathbf{\hat{H}}_{q}^{\text{H}}-{{\mathbf{\hat{H}}}_{q}}{{\mathbf{F}}_{q}}-\mathbf{F}_{q}^{\text{H}}\mathbf{\hat{H}}_{q}^{\text{H}}+\mathbf{I} +\left(\sigma _{n}^{2}+P\sigma _e^2\right) {{\mathbf{F}}_{q}}\mathbf{F}_{q}^{\text{H}} \right).\label{eq22}
\end{align}
After achieving the simplified exact expression of MSE for the $q$-th sensor, by taking the first order derivative of $\text{MSE}_q$ in \eqref{eq22} with respect to $\mathbf{F}_q$ and setting to zero (i.e., $\frac{\partial{\text{MSE}_q}}{\partial{\mathbf{F}_{q}}} = 0$), we arrive at the closed-form solution of the robust precoding matrix as shown in Proposition 1.
\end{IEEEproof}
\begin{remark}
	For the non-robust precoding case (e.g., the precoding design is performed without considering the existence of errors in the estimated channel matrix), under the same derivation procedure, we can obtain the precoding matrix $\mathbf{F}_q^{nr}$ displayed as
	\begin{equation}
		\mathbf{F}_{q}^{nr} = {\left( {{{\mathbf{\hat{H}}}^\text{H}_q}{\mathbf{\hat{H}}}_q +  {{{\sigma _n^2}}} {\mathbf{I}}} \right)^{ - 1}}{{\bf{\hat{H}}}^\text{H}_q}. \label{eq24}
	\end{equation}
\end{remark}

\begin{remark}
Note that our proposed robust precoder can be extended to the fractional Doppler case. In this case, the expression of the estimated $\hat{h}_{p,t-1}$ in \eqref{eq7} needs to be rewritten based on the corresponding channel estimation approach. Since the statistic of estimation error under the fractional Doppler case is very challenging to obtain, for analytical convenience, we assume the channel estimation error to be Gaussian distributed and the similar derivation steps presented in \eqref{eq8} - \eqref{eq22} can be applied to achieve the robust precoding design. The detailed derivation is presented in Appendix A. The effectiveness of the proposed robust precoder for the fractional Doppler case will be demonstrated in the simulation section.
\end{remark}

\section{Simulation Results}
In this section, we evaluate the performance of the proposed robust precoding scheme for the considered system by simulations. We show the results for normalized MSE (NMSE) of the computation as the performance metric, which is defined as the ratio of the MSE in \eqref{eq11} to the mean square of the true value.
Unless otherwise specified, the simulation parameters are set as follows: the number of Doppler bins $N=64$, the number of delay bins {$M=64$}, the number of sensors $Q=6$, the number of independent paths between each sensor and the AP $P=3$. The pilot signal-to-noise ratio (SNR) for channel estimation is set to $\frac{x^2_o}{\sigma^2_w} = 30$ dB. The delay taps and Doppler taps of each path are set to a random integer in $\left[ 0,{{l }_{\max }} \right]$ and $\left[ -{{k }_{\max }},{{k }_{\max }} \right]$, respectively, where ${l}_{\max }=4$ and ${k }_{\max }=2$. Since we assume that each sensor performing the sensing task is separated from each other by sufficiently large distance, the results are generated based on independent channel conditions.

\begin{figure}[t]
	\centering
	{\includegraphics[width=0.6\linewidth]{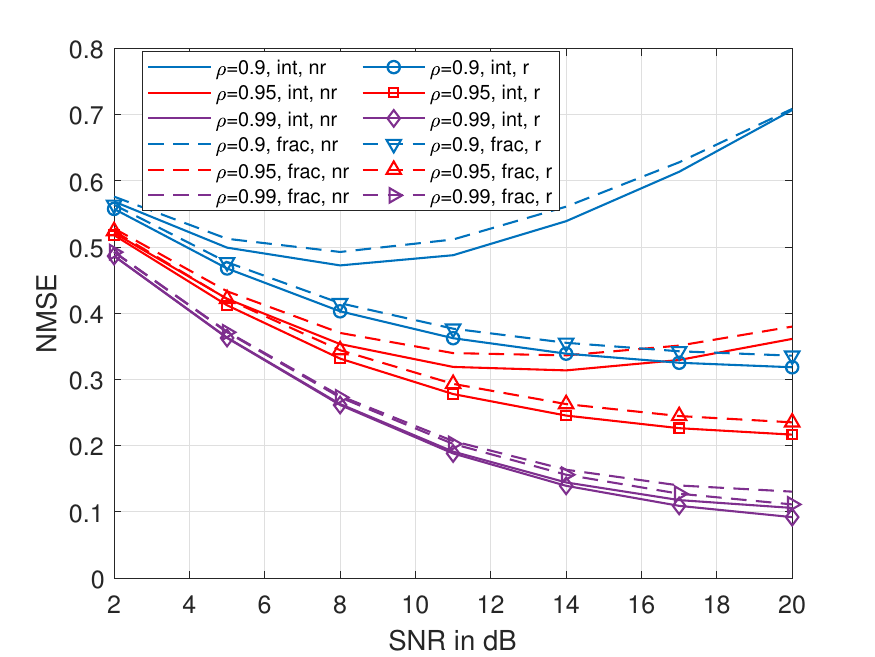}}
	\caption{Computation NMSE versus SNR under different estimation errors for integer Doppler case and fractional Doppler case when delay taps and Doppler taps are accurate.}
	\label{fig3}
\end{figure}

Fig.~\ref{fig3} plots the computation NMSE versus SNR, which is defined as the ratio of the power of each symbol to the noise power at the receiver, under different estimation errors for both integer Doppler case and fractional Doppler case. For the purpose of comparison, the results for the non-robust precoding case are also plotted. Here, the estimation of delay taps and Doppler taps are assumed to be accurate. Fig.~\ref{fig3}, it can be observed that the computation NMSE of the robust precoding is lower compared to the computation NMSE of the non-robust precoding, which is more significant when the estimation error is large. In addition, Fig.~\ref{fig3} shows that, as the increasing of SNR, the computation NMSE under the non-robust precoding design decreases at first and then slightly increases. In other words, the high SNR can deteriorate the computation NMSE under the non-robust precoding design. It can be explained as follows. When the SNR is very small, the noise from the receiver is very large, especially compared to the estimation error, and it plays the dominant role in determining the performance of computation NMSE. Hence, the increase of SNR can reduce the impact of the noise, thereby improving the system performance. However, when the SNR becomes very large, the noise intensity becomes very small compared to the estimation error in the channel matrix calculated from the estimated CSI, which worsens the computation NMSE due to the ignorance of the estimation error. This problem can be solved by using the proposed robust precoding scheme. This is due to the fact that the compensation unit matrix is added to guarantee that the error in the channel matrix calculated from the estimated CSI is not dominant, thus ensuring that the computation NMSE maintains a continuous decreasing trend with the increase of SNR. 
Besides, from Fig.~\ref{fig3}, it can be observed that the proposed robust precoding scheme is effective for the fractional Doppler case. Due to the difficulty in perfectly estimating fractional Doppler taps, there is an extra error in the estimated channel gains in the case of fractional Doppler, together with the Gaussian assumption for the channel estimation error, which consequently results in slightly worse computed NMSE computation compared to the integer Doppler case.

\begin{figure*}[t]
	\centering
	\subfloat[Only delay taps experience offset errors.]  {\includegraphics[width=0.34\linewidth]{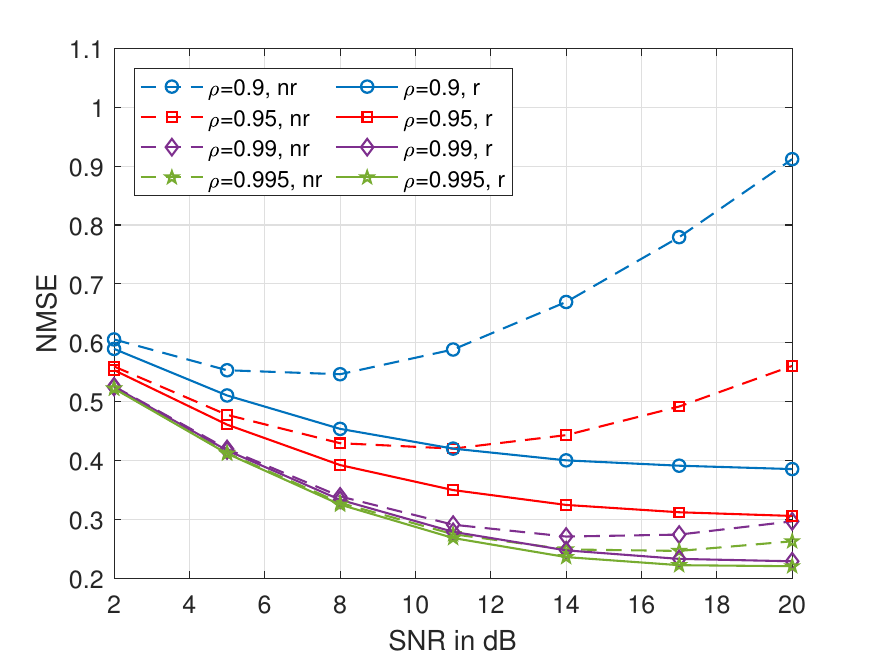}\label{a}}
	\subfloat[{Only Doppler taps experience offset errors.}] {\includegraphics[width=0.34\linewidth]{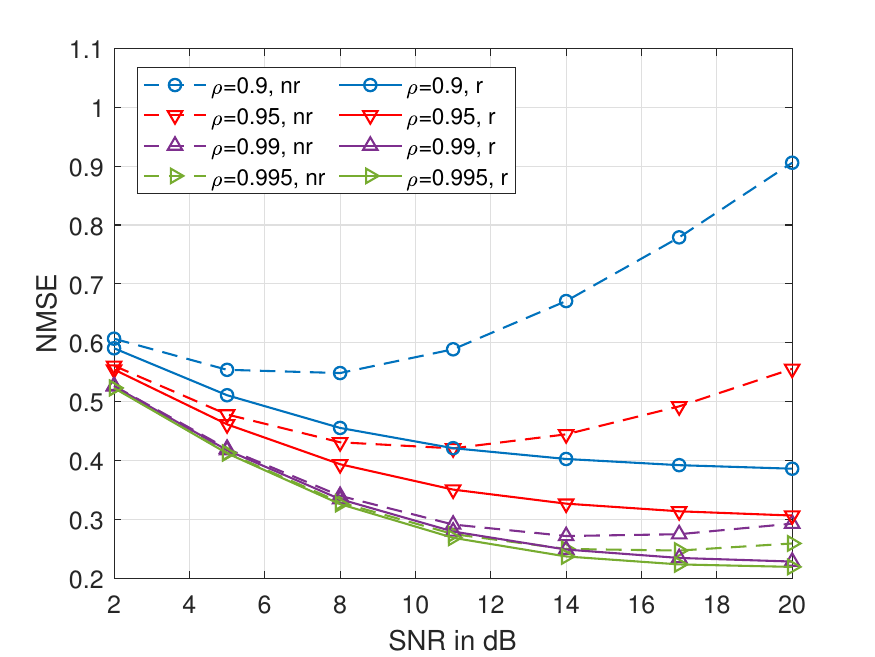}\label{b}}
	\subfloat[{Both delay taps and Doppler taps experience offset errors.}] {\includegraphics[width=0.34\linewidth]{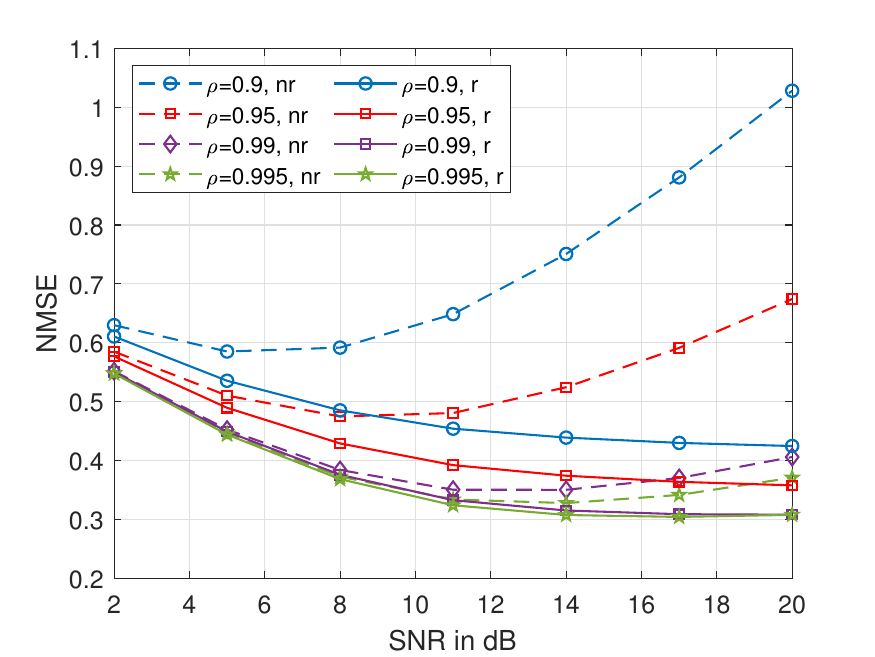}\label{c}}
	\\
	\caption{{Computation NMSE versus SNR under different channel estimation noise in the case of integer Doppler case when delay taps and Doppler taps are imperfect.}}
	\label{Offset_err}
\end{figure*}

Fig.~\ref{Offset_err} plots Computation NMSE versus SNR under different channel estimation noise in the case of integer Doppler case when delay taps and Doppler taps are imperfect, where Fig.~\ref{Offset_err}(a), Fig.~\ref{Offset_err}(b), Fig.~\ref{Offset_err}(c) shows the results where only delay taps experience offset error, only Doppler taps experience error, and both delay taps and Doppler taps experience offset error, respectively. Therein, $1$ grid offset error with $10\%$ probability for delay taps and Doppler taps is assumed. From Fig.~\ref{Offset_err} we can find that in this case, the performance of the non-robust precoding scheme becomes worse in the high SNR scenarios (e.g., the increasing trend of non-robust precoder is much more obvious when compared with Fig.~\ref{fig3} since the extra error of CSI is involved. As for the proposed robust precoding scheme, the computation NMSE is still decreasing with the increment of SNR, i.e., the precoder design can still maintain convergence. This indicates the robustness of our developed precoding design to some extent, especially compared to the non-robust precoding scheme. Moreover, according to Fig.~\ref{Offset_err}, the impact of offset errors in delay taps and Doppler taps on computation NMSE is almost the same, while the joint error effects lead to worse performance degradation.

\begin{figure}[t!]
	\centering{\includegraphics[width=0.6\linewidth]{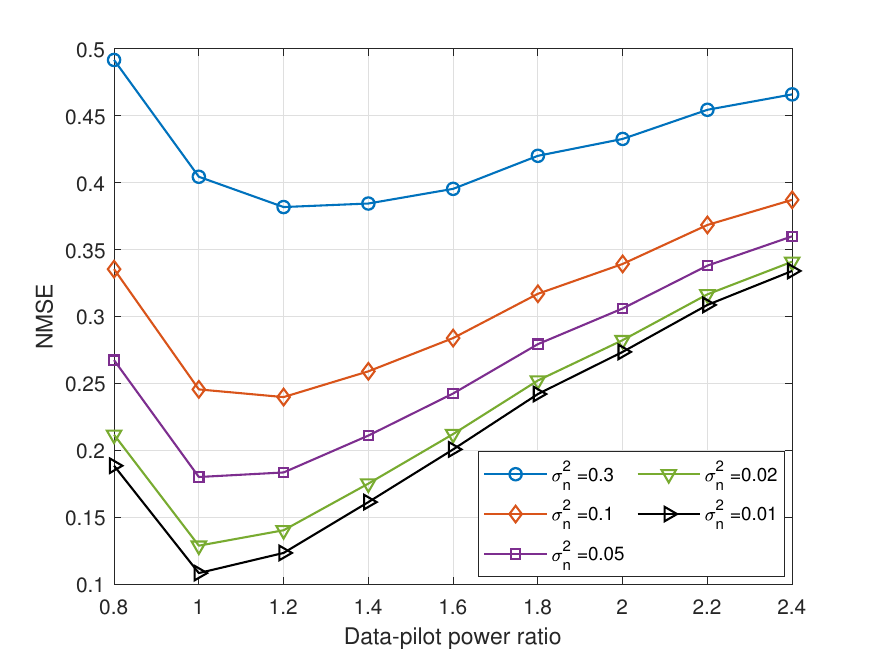}}
	\caption{Computation NMSE versus the ratio of data power and pilot power under different noise levels when $\rho=0.99$.}
	\label{fig5}
\end{figure}
Fig.~\ref{fig5} plots the computation NMSE from the robust precoding scheme versus the ratio of data power and pilot power, under different noise levels when $\rho=0.99$. From Fig.~\ref{fig5}, the computation NMSE drops at first and then rises with the increasing of the power ratio. Under the considered system parameters, the optimal power ratio happens around 1-1.2.  
This is because when the power ratio is small, the data power is so small, e.g., even may be much smaller than the receiver noise power, which consequently results in a large computation error. With the rising of the power ratio, the data symbol power increases, whereby boosting up the computation NMSE. However, a further increase in the power ratio can worsen the system performance. Since the channel estimation error increases as the pilot power decreases. The high enough transmission power for data still cannot compensate the error caused by channel estimation. On the whole, the interplay of these two factors leads to this trend. 

\section{Conclusions}
In this work, we investigated an OTFS-based AirComp system,  where a UAV is deployed to collect the data from a number of dual-functions sensors by AirComp technology. Based on the considered system, a transmission framework without CSI feedback overhead was developed by exploiting the echo from the UAV for channel estimation at the sensor's side. Moreover, the OTFS waveform was adopted to eliminate the effect of the time-frequency dual-selective channel on AirComp.
Then under the consideration of the errors from the noise as well as the outdated CSI, a robust precoding scheme based on the statistical properties of errors was designed. Simulation results show that the proposed robust precoding scheme can effectively reduce the computation MSE, especially in the presence of large channel estimation errors. In addition, a suitable power allocation can also improve the computation accuracy. Our future work can include the power allocation optimization for the estimation and data transmission, and the transmission design with all channel estimation errors considered.  

\section*{Appendix A}
\begin{figure}[htpb]
	\centerline{\includegraphics[width=0.8\linewidth]{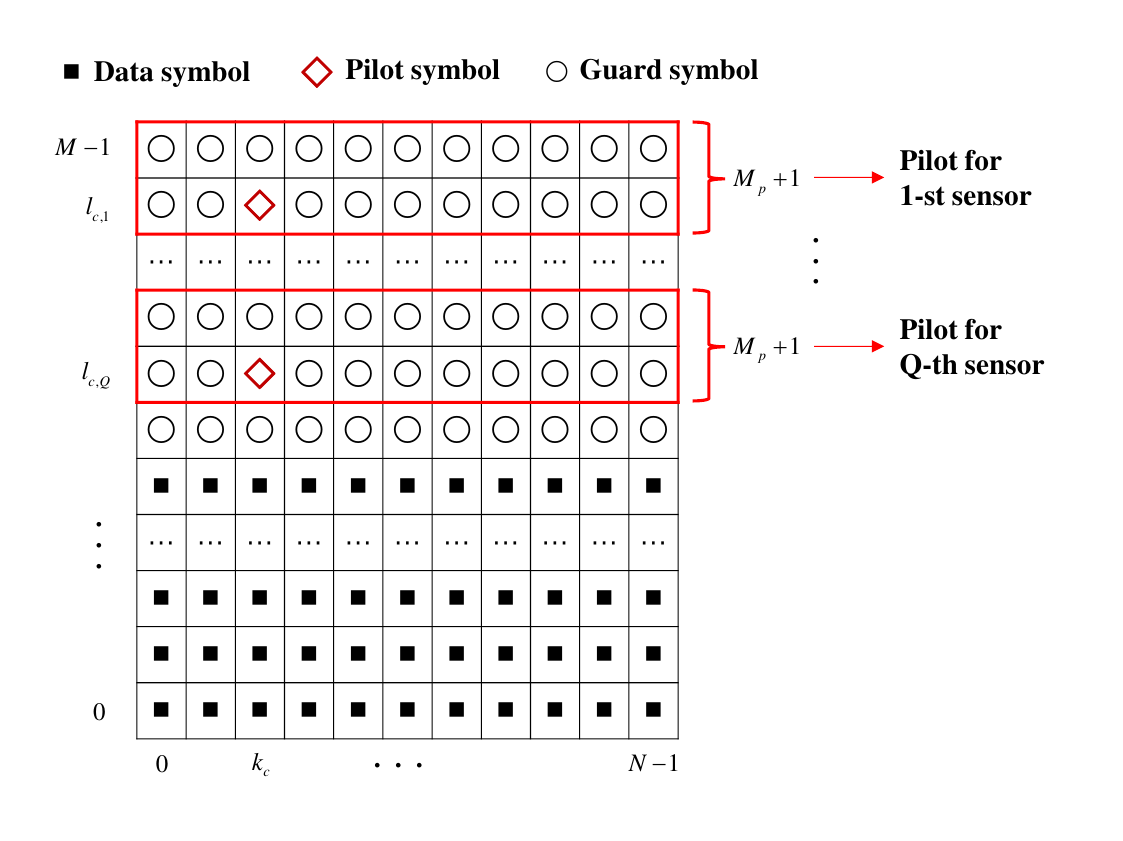}}
	\caption{The schematic diagram of symbol arrangement in the case of fractional Doppler.}
	\label{fig6}
\end{figure}

For the fractional Doppler case, the channel in the delay-Doppler domain is spread across all the Doppler indices due to the dispersion of fractional Doppler. Therefore, the pilot pattern for the sensor node is reconstructed as shown in Fig.~\ref{fig6}. Specifically, for the $q$-th sensor node, one pilot symbol $x_o$ is placed at $\mathbf{X}[l_{c,q},k_c]$ and the guard symbols are arranged at $\mathbf{X}[l,k], l\in [l_{c,q}\leq l\leq l_{c,q}+M_p], k\in\left[ 0\leq k\leq N-1 \right], [l,k]\ne[l_{c,q},k_c]$, i.e., the pilot symbols and guard symbols occupy all the Doppler grids of the corresponding delay grids to avoid interference with data symbols. In the received echo signal $\mathbf{Y}$, the symbols $\mathbf{Y}[l,k]$ for $l\in [l_{c,q}\leq l\leq l_{c,q}+M_p], k\in\left[ 0\leq k\leq N-1 \right]$ will be used for the channel estimation.

Since the pilot symbols of different sensor nodes are arranged orthogonally, there is no mutual interference between the pilot symbols in the received echo signal received at each sensor node. In the following, the channel estimation for the $q$-th sensor is analyzed as an example and the subscript $q$ is omitted.

For the $q$-th sensor node, let $l_p, k_p+\kappa_p, h_p$ denote the channel delay tap, Doppler tap, and channel gain coefficients of the $p$-th path, respectively, where $\kappa_p \in [-0.5,0.5]$ is the fractional Doppler tap. According to \cite{bxx}, the round-trip delay tap and Doppler tap can be expressed as $\tilde{l}_p =2 l_p, \tilde{k}_p +\tilde{\kappa}_p  = 2(k_p+\kappa_p)$, respectively. Let $\mathbf{Y}_{\text{pilot}} \in \mathbb{C}^{M_p \times N}$, $\mathbf{y}_i$ and $\mathbf{y}_i(j)$ denote the pilot symbols used for channel estimation in the received echo, the vector consisting of the elements of the $i$-th row in $\mathbf{Y}_{\text{pilot}}$, and the $j$-th element of $\mathbf{y}_i$, respectively. Due to the signal propagation along the $p$-th path, the pilot symbol $x_o$ will affect some symbols of $\mathbf{Y}_{\text{pilot}}$, which can be represented as 
\begin{align}
	{\mathbf{Y}_{\text{pilot}}}\left( {{{\tilde{l}}}_{p}},k_c+{{{\tilde{k}}}_{p}}-a \right)={{x}_{o}}{{h}_{p}}{{e}^{j\frac{2\pi {{l}_{c}}\left( {{{\tilde{k}}}_{p}}+{{{\tilde{\kappa }}}_{p}} \right)}{MN}}}\beta \left( a,{{{\tilde{\kappa }}}_{p}} \right), k_c+{{{\tilde{k}}}_{p}}-a \in \left[0,N-1\right], \label{eqA2}
\end{align}
where $a$ represents the Doppler index deviation from ${{{\tilde{k}}}_{p}}$ and $\beta \left( a,\kappa  \right)=\frac{1}{N}\frac{{{e}^{j2\pi \left( a+\kappa  \right)}}-1}{{{e}^{j\frac{2\pi \left( a+\kappa  \right)}{N}}}-1}$, which indicates the impact of fractional Doppler dispersion. Based on \eqref{eqA2}, we can obtain the channel estimation for the fractional Doppler case as follows.

The estimated value of the round-trip delay tap is first determined based on the indexes of the  $P$ row vectors with the highest power in $\mathbf{Y}_{\text{pilot}}$. Then, the estimated value of the integer Doppler tap is obtained as the index of the symbol with the highest power in the corresponding row. Subsequently, based on the form of fractional Doppler dispersion, the maximum likelihood estimation method is applied to obtain the optimal fractional Doppler tap from a pre-defined set of fractional Doppler candidates. Algorithm 1 summarizes the ML-based channel estimation method for the fractional Doppler case.

\begin{algorithm}[hpbt]
	\setstretch{1.5}
	\caption{ML-based channel estimation method for the fractional Doppler case.}\label{alg:alg1}
	\begin{algorithmic}[1]
		\STATE {\textbf{Input:}} Received pilot signal $\mathbf{Y}_{\text{pilot}}$, fractional Doppler candidates $\mathbb{K}=\left\{ -0.5,-0.4,...,0.5 \right\}$.
		\STATE {\textbf{Initialize:}} $p = 1$
		
		\STATE {\textbf{while}} $p\le P$ \textbf{do}
		
		\STATE \hspace{0.5cm} Find the index $I_p$ of the $p$-th largest power row in $\mathbf{Y}_{\text{pilot}}$, and let $\widehat{{{{\tilde{l}}}_{p}}} = I_p$
		
		\STATE \hspace{0.5cm}  $\widehat{{{{\tilde{k}}}_{p}}}=\underset{j}{\mathop{\arg \max }}\,\left| {{\mathbf{y}}_{\widehat{{{{\tilde{l}}}_{p}}}}}\left( j \right) \right|-k_c$

		\STATE \hspace{0.5cm}  $ {{\mathbf{s}}^{{{\kappa }}}}\left( n \right)=\frac{{{\mathbf{y}}_{_{\widehat{{{{\tilde{l}}}_{p}}}}}}\left( \widehat{{{{\tilde{k}}}_{p}}} \right)}{\beta \left( 0,{{\kappa }} \right)}\beta \left( {\widehat{{{{\tilde{k}}}_{p}}}}-n,{{\kappa }} \right),\ n=0,1,...,N_p-1,\ \forall \kappa \in \mathbb{K} $
		
		\STATE \hspace{0.5cm} $\widehat{{{{\tilde{\kappa }}}_{p}}}=\underset{\kappa \in \mathbb{K}}{\mathop{\arg \min }}\,\left\| {{\mathbf{s}}^{\kappa }}-{{\mathbf{y}}_{\widehat{{{{\tilde{l}}}_{p}}}}} \right\|_{2}^{2}$
		
		\STATE \hspace{0.5cm} ${{\hat{h}}_{p}}=\frac{{{\mathbf{y}}_{_{\widehat{{{{\tilde{l}}}_{p}}}}}}\left( \widehat{{{{\tilde{k}}}_{p}}} \right)}{{{x}_{o}} \beta \left( 0,{\widehat{\tilde{\kappa }_{p}}} \right) {{e}^{j\left( 2\pi {{l}_{c}}\left( \widehat{{{\tilde{k}}}_{p}}+\widehat{{{\tilde{\kappa }}}_{p}} \right)/\left( MN \right) \right)}}}$

		\STATE \hspace{0.5cm} $\hat{l}_p = \frac{1}{2}\widehat{{{{\tilde{l }}}_{p}}},\   (\hat{k}_p+\hat{\kappa}_p) = \frac{1}{2}\left(\widehat{{{{\tilde{k }}}_{p}}}+\widehat{{{{\tilde{\kappa }}}_{p}}}\right)$
		
		\STATE \hspace{0.5cm} $p=p+1$
		
		\STATE \textbf{end while}
		
		\STATE {\textbf{Output:}} The estimated CSI $\hat{l}_p$, $\hat{k}_p+\hat{\kappa}_p$, $\hat{h}_p\ \left(p=1,2,...P\right)$.
	\end{algorithmic}
	\label{alg1}
\end{algorithm}

Under the proper pilot power setup, the estimate for the delay taps and the integer Doppler taps $\hat{l}_p$ and $\hat{k}_p$ can be considered to be accurate. However, due to the limited size of the fractional Doppler candidate set $\mathbb{K}$, $\hat{\kappa}_p$ is not completely accurate. According to Algorithm 1, $\hat{h}_p$ can be further expressed as 
\begin{align}
	{{{\hat{h}}}_{p}}& =\frac{{{h}_{p}}{{x}_{o}}\beta \left( 0,{{{\tilde{\kappa }}}_{p}} \right){{e}^{j\frac{2\pi {{l}_{c}}}{MN}\left( {{{\tilde{k}}}_{p}}+{{{\tilde{\kappa }}}_{p}} \right)}}+{{w}_{p}}}{{{x}_{o}}\beta \left( 0,\widehat{{{{\tilde{\kappa }}}_{p}}} \right){{e}^{j\frac{2\pi {{l}_{c}}}{MN}\left( {{{\tilde{k}}}_{p}}+\widehat{{{{\tilde{\kappa }}}_{p}}} \right)}}} \nonumber\\ 
	& ={{h}_{p}}\frac{\beta \left( 0,{{{\tilde{\kappa }}}_{p}} \right)}{\beta \left( 0,\widehat{{{{\tilde{\kappa }}}_{p}}} \right)}{{e}^{j\frac{2\pi {{l}_{c}}}{MN}\left( {{{\tilde{\kappa }}}_{p}}-\widehat{{{{\tilde{\kappa }}}_{p}}} \right)}}+\frac{{{w}_{p}}}{{{x}_{o}}\beta \left( 0,\widehat{{{{\tilde{\kappa }}}_{p}}} \right){{e}^{j\frac{2\pi {{l}_{c}}}{MN}\left( {{{\tilde{k}}}_{p}}+\widehat{{{{\tilde{\kappa }}}_{p}}} \right)}}}.  \label{eqA3}
\end{align}

Here, we just present one channel estimation example. Note other channel estimation algorithms can be applied and the corresponding $\hat{h}_p$ can be analyzed accordingly. From \eqref{eqA3}, we can find that in the case of fractional Doppler, it is very challenging to obtain the statistical characteristics of the channel estimation error. Here, for analytical convenience, we approximate the channel estimation error as a complex Gaussian variable with zero mean and variance $\sigma^2_e$, where $\sigma^2_e = \rho^2 \frac{\sigma^2_w}{x^2_o} + \frac{(1-\rho^2)}{P}$ is the same as in the integer Doppler case. The following derivations and the statistical characteristics of the error matrix are almost the same as the counterpart of the integer Doppler case, i.e., the process of \eqref{eq8} to \eqref{eq22}. The only differences are the expression of the time domain channel matrix $\mathbf{H}^{\text{TD}}$ and the error in the time domain matrix in \eqref{eq10c}. In the case of fractional Doppler, $\mathbf{H}^{\text{TD}}$ is expressed in \cite{frac_Doppler} as	 
\begin{equation}
	\mathbf{H}^{\text{TD}} =\sum_{p=1}^{P} h_{p} \mathbf{\Pi}^{l_{p}} \mathbf{\Delta}(p), \label{eqA4}
\end{equation}
where $\mathbf{\Delta }\left( p \right)=diag([{{w}^{0}},{{w}^{1}},\ ...\ ,{{w}^{MN-1-{{l}_{p}}}},{{w}^{-{{l}_{p}}}},\ ...\ ,{{w}^{-1}}])$, $w={{e}^{\frac{j2\pi }{MN}({{k}_{p}+\kappa_p})}}$. The error in the time domain matrix for the fractional Doppler case is correspondingly written as
\begin{equation}
	\mathbf{E}^{\text{TD}}=\sum\limits_{p=1}^{P}{{{e}_{p}}}{{\mathbf{\Pi }}^{{{l}_{p}}}}\mathbf{\Delta }\left( p \right). \label{eqA5}
\end{equation}

%
%

%
%

\vfill

\end{document}